\documentstyle[sprocl,epsf]{article}

\renewcommand{\bar}[1]{\overline{#1}}
\newcommand{\etal}{{\em et al.}}
\newcommand{\ie}{{\it i.e.}}
\newcommand{\eg}{{\it e.g.}}

\begin{document}

\title{CP AND B PHYSICS: PROGRESS AND PROSPECTS}

\author{J. D. Bjorken}

\address{Stanford Linear Accelerator Center \\
Stanford University, Stanford, California 94309}

\maketitle
\abstracts{This summary of the 2nd International Conference on $B$
Physics and CP Violation (Honolulu, 24--27 March, 1997) contains, in
addition to what is implied in the title, some extended remarks on the
limitations of theory as well as speculations regarding nonperturbative
enhancement of decay modes of the class $b \rightarrow s  +$ charmless
hadrons.}

\section{Generalities}

It is a privilege for me, but something of a disadvantage otherwise,
that I give this summary. I have not been active in $B$-physics for
some time. The last time I did review the subject was in the mid 
1980s \cite{ref1}. So, as an outsider, I will neither try to provide a
compendium of the many interesting contributions, nor a comprehensive
summary of the field. There will be few if any detailed quotations of
results or exhibits of plots from the contributors to the meeting, and
this overview will most likely be rather unbalanced.

On the other hand, there are some marginal advantages to being an
outsider. I perhaps can feel the changes which have been occurring
within the field in the last decade more easily than those of you who
are embedded within it full-time. And unlike some of our subfields
which feel all too static, there really is a dynamic change in the
CP/$b$-quark business.  The primary driving force in physics is the
technological progress, next the experiments which benefit from it, and
finally the theory. In all three areas there have been most impressive
advances. At DESY with DORIS/ARGUS, but especially at Cornell with
\hbox{CESR/} CLEO, we have seen the storage ring performance steadily improve
and the detector capability enhanced. CDF, SLD, and the LEP experiments
have discovered $B$-physics and have done much more with it than at
least I anticipated would be done. The asymmetric electron-positron
collider idea emerged and has gone from the R\&D phase to the
construction phase; it of course provides great promise for
CP-violation-level $B$-physics. The new round of kaon-decay experiments
are extraordinary in the statistical power and level of precision being
attained. Charm physics at Fermilab and at CERN in fixed-target mode
has been remarkably productive not only in the data acquired but also
in the technological advances which those experiments pioneered.  The
technological changes most important are the microvertex detectors and
the high rate, large bandwidth data acquisition systems. And the
beautiful photon detection capability at CESR via CsI (and at L3 with
BGO) has made a huge difference. And at long last the ring-imaging
Cerenkov technique seems to have matured.

At present there is a very big data base of high quality. The CKM
parameters are already measured quite well, although we all hunger for
still more accuracy. And on the theoretical side, there has emerged
what I used to call Wisgur, in honor (???) of the originators Wise and
Isgur \cite{ref2}, and which is now called Heavy Quark Effective Theory
(HQET). This development transcended the model-building approach which
preceded it, and provides a solid foundation for analysis of
semileptonic decays and is an important tool for a variety of other
problems.

And the community has grown into a very large one, with very high
professional standards. I hope that what I perceive is not just the
progress in producing transparencies, so impressive that it becomes
impossible to doubt the truth of what is displayed upon them.

Ten years ago the big question was whether CP violation in the $B$
system could ever be seen \cite{ref3}. Now the question has simply
become who will see it first. The race is on, and my (conservative)
guess of completion time is no later than 2001 or so. Coming in first
is rightly important to the participants in the race. But it has to be
emphasized that the end of the race is just the beginning of the
field---the first measurements must be supplemented by many others to
get the big picture of what any one measurement really means.  So
losers really won't lose. Everyone should win, because the variety of
CP violating phenomena which becomes accessible at a level of
difficulty not far beyond the discovery mode is very large. There is
plenty of room for everybody.

Nowadays, the issues in the $B$ business seem to have become sharply
defined. One sees in the projections of new initiatives quite precise
specifications of what must be accomplished, \eg\ the precision to
which $\sin 2 \beta$ or $x_s$ will be measured, etc. \cite{ref4}  It
looks almost like engineering. Is it really true that the future of
$B$-physics will be another round of LEP I Yellow-book physics, with
many fine results of exquisite precision, but constrained within
standard-model expectations? Of course nobody knows. This could happen,
and it often \hbox{seems} that this is the standard presumption. On the
other hand there might be quite strong shocks to the standard picture. 
In particular, one might ask

\begin{enumerate}
\item
Is the unitarity triangle not? 

\par\noindent
or
\item
Is the unitarity triangle not enough? 

\par\noindent
or, on the other hand,
\item   
Is the unitarity triangle right? 
\end{enumerate}

\noindent
Now for the explanations of what is meant, in turn: 

\begin{itemize}

\item
Maybe the unitarity triangle is not an interesting one, \ie\ it is
degenerate. Perhaps it will be discovered that there exists an
unexpected discrete symmetry, like the occasional discrete symmetries
entertained for reasons of expediency by SUSY model-builders and
others, but this one immediately accessible. This one could be called
electroweak $T$ conservation, \ie\ there does exist a basis where all
CKM parameters are real. This is what is implied by the superweak
theory and also by the aspon model, presented here by Paul Frampton
\cite{ref5}. The data as presented at this meeting by Roger
Forty \cite{ref6} is only marginally consistent with this hypothesis,
but it was emphasized by many that the systematic errors are dominated
by theoretical uncertainty and cannot be treated reliably by standard
statistical methods. I agree with this conventional wisdom. And since
the main culprit in error production is $V_{td}$, its measurement by
$B_d$ and $B_s$ mixing (or $K \rightarrow \pi\nu\bar\nu$) seems to be
the best way of improving the situation. But imagine the impact of the
electroweak $T$-conserving scenario: the kaon experiments see no
evidence for nonvanishing $\epsilon^\prime$, and everyone measures
everything at the $B$-factories, with no trace of CP violation.
Fantastically exciting!! It's the arrival of new, unexpected physics,
not with a bang but with the whimpers of a thousand experimentalists.

\item
Maybe the CKM parametrization of CP violating effects is not enough,
and that in addition to the standard sources of CP violation  there are
extra contributions. This could lead to inconsistent ``determinations"
of the angles $\alpha$, $\beta$, and $\gamma$. It could also lead to CP
violation occurring where it is not expected to occur, such as in
$\psi\eta$, $\psi\phi$, $b \rightarrow s \gamma$, etc. One moral to the
story is to look everywhere for CP violation, even in places shunned by
respectable theorists. Another moral is that it is important to measure
the standard CKM parameters as redundantly as possible. Lincoln
Wolfenstein seemed in his talk \cite{ref7} to be somewhat distressed by
the difficulties in sorting things out in this scenario. I can only
wish that this were our problem.

\item
Maybe there is a right angle in the unitarity triangle, in particular
maybe $\gamma = 90^\circ$. This is not in the same class of dramatic
surprises as the previous two categories, but nevertheless  an observed
regularity of shape of the unitarity triangle might send a rather
strong message. There is a small, elite right-angle club, consisting to
the best of my knowledge of Berthold Stech and myself. Harald Fritsch
qualifies as a corresponding member (e-mail only), having also
advocated a right angle \cite{ref8}, but the wrong one ($\alpha$).  The
idea is Stech's \cite{ref9} and is based on the hypothesis that, in a
basis where the up-quark mass matrix is diagonal, the down quark mass
matrix has real diagonal elements and small imaginary-antisymmetric
off-diagonal elements \cite{ref10}. The smallness requirement is that
second-order perturbation theory suffices in the calculation of CKM
parameters. So it was a delight to witness the exhibit of the unitarity
triangle by Forty\cite{ref6}, so consistent with Stech's hypothesis,
even though the consistency is without much meaning (remember item 1
above!).

\end{itemize}

\section{Trouble List}

To me, it is more important to pay attention to the difficulties in the
subject, rather than the successes---even in a summary like this. It is
in thinking about the difficulties more than in celebrating the
progress that advances are made. In this meeting, a variety of problems
were mentioned. None are anywhere near a crisis stage, but many are
worth some serious attention. What follows is, first, a trouble list
that I have put together, but only from my limited exposure to the
subject at this meeting, with very brief descriptive remarks and
references to the source material. This will be followed by some
elaborations on the first five items on the list:

\begin{itemize} 
\item{Limitations of the theory:} 

Already in the discussion of the unitarity triangle above \cite{ref6},
we have encountered the regrettable fact that theoretical, not
experimental errors dominate. Lattice QCD provides valuable assistance
on some problems, but its range of applicability is rather limited. And
while HQET provides an excellent framework \cite{ref11} for some of the
theoretical questions, this is not true in general. Furthermore, the
applicability of perturbative, short-distance, weak-coupling QCD is in
my mind at best marginally defensible. More will be said about this
later.

\item{Smallness of the semileptonic branching ratio and
charm-counting:}

The value of the semileptonic branching ratio has for a long time been
a bit small for theory, suggesting an enhancement of some of the
nonleptonic modes \cite{ref12,ref13}. The most attractive hypothesis is
that it is the $c \bar c s$ class of final states that are enhanced,
due to the small $Q$-value and the possibility of resonances, etc. as
the enhancement mechanism \cite{ref12}. This would imply a
larger-than-expected number of charm hadrons produced per decay. But
not enough is seen to satisfy the proponents \cite{ref57}.

\item{Magnitude of the $b\rightarrow s +$ charmless channels:} 

The charm-fraction problem has increased the focus on this class of
channels, which are ``gluonic penguin" and/or $c \bar c$ final-state
annihilation channels.  Standard PQCD estimates \cite{ref16} put this
branching ratio in the 1--2\% range. For reasons of new-physics, Kagan
would like an order of magnitude larger \cite{ref13}. And for me 
(as well as others \cite{ref12}) it is
an interesting question whether nonperturbative QCD effects could yield
a larger value. As emphasized by Kagan and Dunietz \cite{ref12,ref13},
measurement of the yield of energetic direct kaons (those not from
charm decays) is a good way to get at this issue experimentally.

\item{Why so many $\eta^\prime$'s in $B$ decays?}

CLEO has presented \cite{ref19} a variety of decay modes featuring
$\eta^\prime$'s. The branching ratio into $\eta^\prime K$ is about $7.5
\times 10^{-5}$, four times larger than the estimate of BSW from theory
\cite{ref20}. The inclusive yield into $K \eta^\prime$ and any number
of additional pions, with the requirement that the $\eta^\prime$ lab
momentum exceed 2 GeV, is an order of magnitude larger. What is the
message, if any, from these new measurements? \cite{refaaa}

\item{Why is the $\Lambda_b$ lifetime so short?}

The $\Lambda_b$ lifetime appears to be only about 80\%\ of the
$B$-meson lifetimes \cite{ref21}.  Short-distance, PQCD-based theory,
according to Neubert \cite{ref11}, will have a very hard time getting
anything less than 90\%\ for the ratio. At present, the measurements of
semileptonic branching ratio of the $\Lambda_b$ give a lower number
than for the mesons.  The analysis \cite{ref43} is difficult, but does
suggest  that the increase in $\Lambda_b$ width
originates in the nonleptonic-decay sector. The experimental numbers
are still semi-soft, but hard enough to create some concern.

\item{Why do the parameters $a_1$ and $a_2$, which describe the
relative importance of the so-called ``color-allowed" and
``color-suppressed" nonleptonic-decay four-fermion interaction terms,
change their relative sign in going from charm decays (opposite sign) to
bottom decays (same sign)?} \cite{ref15}

Berthold Stech \cite{ref22} expressed puzzlement on this problem; this
again may be a signal that thinking that extends beyond the
perturbative-QCD limit is required.

\item{Why is the (now well-observed) \cite{ref23} width of $B$ into
$K\pi$ larger than that into $\pi\pi$?}

The former contribution is in the penguin-induced category, while the
latter can be estimated via factorization and knowledge of charmless
semileptonic decays into $\pi \ell \nu$.

\item{Why is the hyperfine mass-splitting between $\Sigma_b^*$ and
$\Sigma_b$ so large?}

Heavy quark effective theory predicts \cite{ref24} this splitting with
confidence in terms of the corresponding hyperfine interval in the charm
system. The central value emergent from the data \cite{ref25} is more
than a factor of two too large. However, it is still possible to blame the
data, not theory.

\item{What is the phenomenology of the nonresonant semileptonic
decays of $B$ into $(D,D^*) \pi \ell \nu$?}

The bookkeeping on the semileptonic $B$ decays now is good enough to
determine that after $D$, $D^*$, and $D^{**}$ final states are
accounted for, there is still a couple of percent of branching ratio
into the aforementioned channels \cite{ref26,ref27,ref17,ref28,ref11}.
There is at present not good theoretical control on what the physics
underlying those decay modes represents. Since these semileptonic
decays are such a fundamental HQET playground, this represents a rather
important challenge.

\item{What is one to do about ``penguin pollution"?}

There was quite a lot of discussion \cite{ref29} of the problems
created by interference of various CP-violating penguin contributions
with the more calculable ones, introducing difficult-to-anticipate
strong-interaction phases into the analysis, and beclouding the
extraction of accurate values of the CKM phases. This question to me is
easy to answer. First, we should be so fortunate to have that problem.
Experimentalists, please go out and observe these dirty CP violating
processes, without inhibition!! (I hope and trust that this is totally
unnecessary rhetoric.) Then let the further developments be
data-driven. I fully expect that in many such cases the effects of the
pollutants can and will be controlled by ideas on and measurements of
related phenomena which are hard to anticipate in advance. There is
nothing like some data to clear the mind.

\end{itemize}

\bigskip
\noindent
{\bf A few comments on the limitations of the theory:}

There were presented during the meeting a large number of
theoretical contributions based on the techniques of short-distance,
perturbative QCD. These go under the names operator-product expansion,
PQCD, parton-hadron duality, etc.  At the most basic level, they
reflect the choice to describe the phenomenology at the parton level of
quarks and gluons.

It is not self-evident that this is appropriate. There are certainly
successful applications of short-distance methods at energy scales less
than 5 GeV. The total widths of charmonia come to mind, as well as the
elegant analysis of tau decays to extract a value of the strong
coupling constant. Also the whole subject of QCD sum rules resides in
this energy range, although this technique involves art as well as
science. However, success is far from guaranteed. During the meeting
some of the theoretical discussion seemed to imply that one would be
finding parton jets in the final states of $B$ decays. This will not
happen. Electron-positron annihilation at 5 GeV no doubt is initiated
by a quark anti-quark pair, but their hadronization leaves little trace
of jettiness. True, the discovery experiment of ``jettiness" did occur
\cite{ref31} at just this cms energy. But the evidence for jets was
confined to the properties of the leading particle and was subtle at
best. To good approximation at this energy the final state particle
distribution is just phase space.  For nonleptonic $B$ decays,
originated by three, not two, partons, the situation has to be even
more like phase space.

Then there is the question of ``parton-hadron duality", the idea that
on average the hadron-level behavior follows the parton-level behavior.
An important application of this concept is to the important problem of
decay widths. Again we can compare with the situation in
electron-positron annihilation. If one compares the estimate of
hadronic vacuum polarization, say, for spacelike virtual photons, to
the perturbative estimate from quark loops and their radiative
corrections, one gets quite good agreement. If one looks at the decay
widths of the timelike photons, expressed in the famous parameter $R$,
one must be more careful. In the entire range from zero energy up to
5.2 GeV, the estimate of $R$ from local parton-hadron duality is only
reasonably reliable in the interval from 1.7 to 3.0 GeV and from 4.6 to
5.2 GeV. Elsewhere there are resonances which locally distort the
expected behavior by a factor of order unity (per relevant quark
flavor). In the timelike region, one must average over an energy
interval \cite{ref32} to retrieve the short-distance results;  this is
just the uncertainty principle at work. In particular the energy
interval should be large in comparison to the level spacing, which
typically is of order a few hundred MeV.

In semileptonic $B$ decays there is precious little phase space
available, and three discrete levels dominate the final state. So there
is a question of why HQET should work at all. The reason is that the
best applications are independent of the assumption of parton-hadron
duality, and that the physics is very simple. While one might expect
corrections of the form
\begin{equation}
\sim \left(\frac{\Lambda_{\rm QCD}}{m_b-m_c}\right)^n
\end{equation}
these do not appear \cite{ref11}, presumably because the underlying
physics is so simple.

But in the nonleptonic decays, even for the $c \bar u d$ sector, the
physics is less simple. Superposed (in space-time) upon the physics of
the gentle, simple $b\rightarrow c$ semileptonic transition, is the
physics of the hadronization of an outgoing $\bar u-d$ pair. If the
pair go out in the same direction, they act as a small color dipole and
do not create much disturbance \cite{ref33}. (This is now checked rather
well experimentally \cite{ref15}.) If they go out in opposite directions,
the more common case, then the color flux-tube that forms between them
sits on top of the $b/c$ system plus light spectator-quark, and one
cannot expect simplicity. So the hypothesis of ``factorization" ,which
essentially implies that these two subsystems do not talk to each
other, becomes suspect. But even assuming factorization, the $\bar u d$
pair would have to have a mass large compared with 1.6 GeV before the
local parton-hadron duality assumption becomes reliable, if one goes by
the indications from electron-positron annihilation data.

So the bottom line is that even for estimating total nonleptonic
widths, HQET may become HQIT, heavy-quark ineffective theory, and that
one may have to take recourse to more explicit models. Many approaches
deserve to be tried, and skepticism about any and all of them must be
kept in the forefront.

There may be a larger role in $B$-decay phenomenology for LQET, light
quark effective theory, than is in place at present. What is meant here
is the long-distance limit of QCD expressed by the effective chiral
action, based on constituent-quark and pion degrees of freedom (plus
eventually a little bit of gluon \cite{ref34}).  This is the formalism
used for (but not limited to) chiral perturbation theory. A great deal
of theoretical progress in this area has been made, both at a
fundamental level and in its applications. There is a school of thought
that the LQET effective action, as catalogued by Gasser and Leutwyler
\cite{ref35}, is already, if not actually derived, well-estimated from
the first-principles QCD Lagrangian by integrating out the gluonic
degrees of freedom, which at intermediate distances are assumed to be
dominated by instanton contributions\cite{ref36}. In particular,
chiral symmetry breaking via this mechanism is claimed to occur
\cite{ref37} and be well understood, and many long-distance properties
of hadrons well-estimated.

\bigskip
\noindent
{\bf A few comments on the experimental trouble list:}

\begin{figure}[htb]
\begin{center}
\leavevmode
\epsfbox{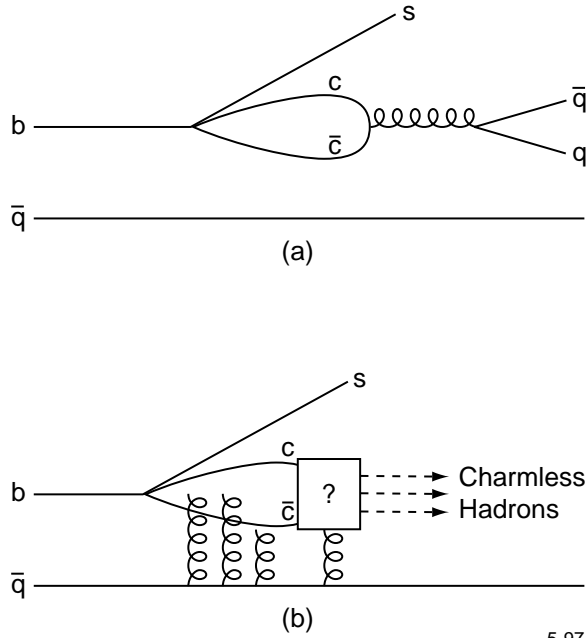}
\end{center}
\caption[*]{Hadronic penguin contributions to $B$-decay: (a)
short-distance, and (b) long-distance.  Time runs from left to right.}
\label{fig1}
\end{figure}

As catalogued above in items 2 to 5, there is a collection of
bookkeeping issues which would be eased if the channels $b \rightarrow
s +$ charmless hadrons were to be ``surprisingly large", \eg\ with
branching ratio 5-15\%\ instead of 1-2\%. Here, I address only the
question of whether there can be enough uncertainty in the
strong-interaction contribution to allow this to happen. (I am no
expert here; in the corridors I found a rather broad spectrum of
opinions on this from theorists more expert than me.) The obvious place
to look is in the region of possible nonperturbative effects. The
gluonic penguin contribution which topologically gives rise to these
final states is shown in Fig. (\ref{fig1}). The penguin is recumbent
for a reason. The $c \bar c$ loop, which dominates the contribution ($u
\bar u$ is very small; $t\bar  t$ simply provides the GIM cutoff at
high mass scales), can have an absorptive part; $c\bar c$ occurs in
real intermediate states, and the issue is how often it might
annihilate into light quarks as shown. If it is going to do it in a big
way, it has to be done more imaginatively than via single gluon
annihilation.

\vspace{.5cm}

If there is an enhancement of the annihilation, it is probable that it
does occur for low $c\bar c$ virtual masses. The perturbative
calculation should work at much higher masses, and common sense tells
us that the nonperturbative effects should be largest when the charm
system moves slowly. One nonperturbative feature of this low mass
system which is reasonably well-established is that its color quantum
numbers fluctuate rapidly. This picture is the basis of the
``color-evaporation" mechanism of onium production \cite{ref38}, as
well as a good candidate interpretation of rapidity-gap phenomenology
in HERA electroproduction studies \cite{ref39}. So although the $c \bar
c$ may be initially color-octet, the probability at later times it is
color singlet can be taken to be 1/9. This mechanism probably is
implemented via soft Coulomb-gluon exchanges with its local color
environment (\ie\ the spectator and $s$ quarks).

In the color-evaporation calculations \cite{ref38}, when the $\bar c c$
has a perturbative mass less than $D\bar D$ threshold, 1/9 of the
events have the final state assigned to onium production. All the rest
are assigned to open charm. This hypothesis agrees with
hadroproduction, photoproduction, and $Z$-decay data on direct
charmonium production very well. However, that agreement is insensitive
to the question of how much of the subthreshold $c\bar c$ production
emerges into open charm; perhaps some considerable fraction finds its
way into charmless final states via annihilation. It is not so easy to
come up with a mechanism for doing this, but we try by grasping at a
straw, namely the large amounts of $\eta^\prime$ seen by CLEO
\cite{ref19}, evocative of glueballs and/or instanton effects helping
things along. If that were to be a clue, then lots of $\eta^\prime$'s
might be also seen in the final states of onium decays. With this
stimulation (due to conversations with Steve Olsen), I checked out the
PDG listings and found nothing in $\psi$ decays, but three remarkably
large exclusive decay channels in $\eta_c$ decays: $\eta\pi\pi$,
$\eta^\prime \pi\pi$, and $K\bar K\pi$. These three channels add up to
about 15\%\ of the total width. They all are obtainable from a
six-fermion, instanton-induced effective vertex $E\cdot B (\bar uu\,
\bar dd\, \bar ss$), or even an 8 fermion vertex with $c\bar c$
appended. All of this seems evocative of communication of the
pseudoscalar $\eta_c$ with $E\cdot B$ which communicates with the QCD
axial anomaly, which communicates with instantons, all of which
communicate with $\eta^\prime$'s as well as pseudoscalar glueballs. I
am told that something like this is advocated by Berkelman, who
emphasizes $\eta_c-\eta^\prime$ mixing. In any case, I myself am
interested in knowing more about those $\eta_c$ decay modes. Are they
already studied and understood by someone?

So maybe the subthreshold, virtual  $c\bar c$ systems also somehow
communicate with these systems, leading with enhanced probability to
charmless final states. As pointed out to me after my talk by a host of
critics, this will have to be done in such a way as not to enhance the
amount of $\eta_c$ in the $B$-decay final states. This, from
experiment, is no larger than 1\%, and is said to be not out of line
with conventional estimations.

Yes, all of this is pretty speculative, and maybe not warranted by the
evidence that something is amiss. But it does lead to a few lines of
study. One, independent of most details, is to check whether
$\Lambda_b$ decays exhibit even more proclivity toward final states
containing eta-primes than the $B$ mesons. If spectator effects matter
in this business, the diquark spectator system for the $\Lambda_b$ may
have a larger influence in driving a nonperturbative mechanism than the
single quark spectator in the $B$. Anyway, perhaps a search for
$\Lambda_b \rightarrow \Lambda \eta^\prime$ is feasable. And, as stated
earlier, the best way to normalize $b \rightarrow s +$ charmless is to
measure the inclusive yield of $B \rightarrow K +$ anything when the
$K$ momentum is large and when the $K$ does not come from a
secondary-charm vertex. Also, for those who consider all of this as
wallowing in minutiae, it should be remembered \cite{ref61} that if
there really is enhancement in such modes, which in $B_s$ decays are $s
\bar s (c\bar  c)$ systems, then this can imply nonperturbative
enhancement of $B_s$ mixing, in particular width-mixing. This in turn
might open up new opportunities for CP-violation studies.

\section{Progress}

What follows will be personal reactions to what feels like recent
progress. Since my perspective is based on a time scale of several
years, it does not necessarily represent what has happened in the last
year, half-year, month, or whatever the correct time quantum is that
separates $B$-physics meetings like this one nowadays. The style will
also be telegraphic; the reader is exhorted to go to the original
sources for the real story:

\begin{enumerate}
\item
Heavy-quark effective theory is working very well in the arena for
which it was invented \cite{ref11}, semileptonic $B$-decays, and the most
important output is a precise determination of $V_{cb}$. There are many
other reliable predictions of this method \cite{ref11}, which, whatever
its limits of applicability, provides a solid rock upon which to build
a large body of theory.

\item
CDF has made an impressive entry into the field \cite{ref42,ref43},
especially with respect to the high-quality determinations of decay
modes containing a $\psi$, and the measurements of lifetimes. It bodes
well for a very productive program in the next collider run at Fermilab.

\item
Lifetime determinations of $B$-mesons and baryons are impressively
accurate \cite{ref43}, so much so that the subfield enters into the study
of small lifetime differences.

\item
The observation of the decay process $b \rightarrow s +\gamma$ is a
remarkable experimental {\em tour de force} by itself \cite{ref44}, and
truly remarkable in attaining an accuracy that one can begin to look
for new-physics effects at the 20\%\ level.

\item
The experimental efforts on $b \rightarrow s + \gamma$ have been
complemented--and complimented--by an extraordinary theoretical effort
\cite{ref45}, with two and three loop effects worked out. This is an
arena where short-distance effects can with reasonable confidence be
expected to prevail---although this is not fully guaranteed. But there
now exists an impressive balance between the accuracy of the
theoretical and experimental numbers, which are somewhere in between
reasonable and marginal agreement. It may happen that, with a modest
increase in accuracy on each side, the process may be elevated onto the
trouble list. If so it will clearly be of enormous importance and
interest.

\item
Measurement of the charmless semileptonic decays is another
experimental {\it tour de force} \cite{ref46}, yielding a rather decent
number for $V_{ub}$. The experiments now are 1$\frac{1}{2}$
generation, with exclusive channels being resolved and the beginnings
of form-factor behavior being probed. The accuracy of $V_{ub}$ seems to
me reasonably adequate already for the immediate future; it is $V_{td}$
which would be nice to pin down more accurately.

\item
The theory of the charmless semileptonic decays is maddeningly
difficult. But I was very much impressed with a tasteful and serious
attack by Burdman \cite{ref47}, using the old-fashioned but powerful
and appropriate dispersion-relation methodology. While so far the
conclusions are modest, I hold out hope that another round along these
lines will yield further benefits.

\item
The most direct way of getting a better $V_{td}$ is in the measurement
of $B_d$ and $B_s$ mixing. Most impressive are the LEP data and
analysis technique\cite{ref48}, which are already at the edge of real
observation of the mass difference at the value anticipated by theory,
namely $x_s$ of order 10--40. I think that the level of accuracy
attained exceeds what many anticipated in advance, thanks in part to
the use of jet charge tagging and inclusive vertexing, as well as
silicon detectors of improved accuracy. However, it appears that one
will need to go beyond LEP to nail down the value of $x_s$.

\item
There has been steady progress in the spectroscopy of the excited
mesonic and baryonic $p$-wave states \cite{ref25}---and perhaps higher.
This is important to do for the charm states as well as bottom states;
HQET is incisively tested by the comparisons, and the higher statistics
for charm may allow finding the higher states more easily and
determining the dynamics well enough that sharp predictions can be made
for the $B$ and $\Lambda_b$ systems. I think it very important that the
spectrum be fully filled in beyond the $p$-wave states and into the
radial excitations and $d$-waves. Not only does this provide very
important information for light-quark QCD spectroscopists, but it can
also refine the single-$b$ flavor-tagging method $(B^{**} \rightarrow
B\, \pi^\pm)$, so important for CP violation studies. In this regard it
was especially interesting to see a candidate for the radial excitation
of the $B$ reported\cite{ref25}.

\item
The two-body rare decays of $B$ into $K\pi$ and $\pi\pi$, which seemed
so hopelessly remote ten years ago, are now being seen \cite{ref23}, with
the two modes separated and with not uninteresting results, as already
mentioned in the trouble list. There are many who eagerly await the
not-yet-reported results on the ratio of $K^+ \pi^-$ to $K^- \pi^+$
yields at CLEO.

\item
The new results from CLEO on $B \rightarrow \eta^\prime K$ and
$\eta^\prime K$ inclusive \cite{ref19} are a fine experimental
achievement, and also may represent an interesting issue for the
theory: are these yields really ``surprisingly large"?

\item
The $D-\bar D$ mixing limits from Fermilab fixed-target charm
experiments\cite{ref50} are remarkably precise. While standard-model
expectations are well below the data, these still probe the new-physics
frontier in interesting ways \cite{ref51,ref52}.

\item
The $B_c$ candidate from ALEPH arrives ahead of schedule \cite{ref25},
but is welcome nevertheless. Now and then such things do arrive ahead
of schedule; in any case this event is a harbinger of things to come. I
also look forward to the first observation of doubly charmed baryons.

\item
As already mentioned, the unitarity triangle exhibited by Forty
\cite{ref6} was most impressive in its purported accuracy and internal
consistency. If only the theoretical uncertainties were less!!

\item
Reports from the lattice world \cite{ref53} showed steady improvement
in the accuracy of the old measurements and the scope of lattice
measurements attempted. However, the quenched approximation, together
with the difficulties in implementing the approximate chiral symmetry
in QCD,  remains a serious obstacle, and any and all attempts to
overcome it are extremely welcome.

\item
Progress reports from the new generation of kaon CP
experiments\cite{ref54,ref55,ref56} bode well for a decisive conclusion
from that long and extremely demanding program. Perhaps the first solid
evidence for CP violation originating from the CKM mixing will emerge
from this source. 
\end{enumerate}

\noindent
In addition to these results, there has been much progress in
experimental technique. I rank the high-rate charm experiments
E791/E687 from Fermilab as very important, especially since they pave
the way for similar experiments at the hadron colliders. And in the
realm of analysis technique, there were impressive new methods
exhibited. The use at CLEO of neural-net methods \cite{ref44} may not
be the very first application, but it seems to me the first time a
claim of a major experimental result is largely based upon that
technique. Also, the use of Fisher discriminants and multidimensional
likelihood analyses, etc., again at
Cornell \cite{ref19,ref23}, in the determination of the $B$ branching
ratios into $K \pi$ and $\pi \pi$ appears to me to be another
escalation in statistical sophistication.

Another landmark analysis has come from ALEPH, in the introduction of
the ``amplitude method" for the interpretation of $B_s$ mixing data
\cite{ref48}. This technique is already becoming a standard for the
field. I am not sure that the quality of the present limits on $B_s$
mixing derives directly from this analysis method, but if nothing else
it creates clarity in assessing the experimental situation.

On the theoretical side, the compendium of suggestions for CP-violation
observations continues to grow. A list culled from the talks, no doubt
incomplete, includes the following:

\begin{enumerate}
\item
Rate asymmetries in the $B$ decays to $K +$ charmless nonleptonic final
states, either in exclusive channels or even inclusively
\cite{ref58,ref13}.

\item
CP violation in $B \rightarrow \phi K_s$, a special case of the 
above \cite{ref59}.

\item
Rate asymmetries \cite{ref51} in $D \rightarrow \rho\pi$.

\item
Rate asymmetries \cite{ref60} in $B \rightarrow D^0 K^-$, with the $D^0
\rightarrow K \pi$ of either \hbox{strangeness}.

\item
CP violation studies which exploit the difference in lifetimes of the
two $B_s$ mass eigenstates \cite{ref61}.     
\end{enumerate}

Again, I exhort, I hope unnecessarily, experimentalists to check
everywhere for CP violating effects. I can imagine a situation where
the first observation occurs in a subregion of a Dalitz plot for a
messy final state (of large branching ratio), where strong final-state
phase shifts conspire to create a maximal effect, much to the surprise
of conservative theorists. Lucky maximal effects might after all
provide quite good measurements of CKM CP-violating parameters via the
setting of lower bounds. Theory might end up being data-driven, rather
than the data being theory-driven.

\section{Prospects}

The present progress derives from technological advances. And the
technological revolution will continue, with no saturation in sight.
The new electron-positron machines are promised to operate at
luminosities of $10^{33}$--$10^{34}$, values that seemed ten years ago
most unreasonable. This will not be easy, and new buzzwords like PEI
(photoelectric instability) and FII (fast ion instability) appear in
the talks \cite{ref62} as sobering reminders of the difficulties. On
the hadron-hadron front, at the LHC \cite{ref63} (and perhaps with a
new opportunity at C0 at Fermilab \cite{ref64}) there should be for the
first time at least one dedicated $B$-physics detector existing in
hadron-collider mode. The enormous yield of $b$'s and $c$'s produced at
hadron colliders will create opportunities which are off-scale relative
to what exists in present data.

In order to exploit the high rates and luminosities, there should---and
almost certainly will be---major advances in detection capability. The
emergence of micropixel detectors of silicon or diamond \cite{ref65}
will, in my opinion,  be a technological advance comparable to that
created by the silicon microstrips. The high rates and bandwidth
attained in the Fermilab charm experiments will be augmented even more
in the next-generation detectors. I found it amusing to hear Brad Cox
describe the LHC-B plan to run at luminosities of $2\times 10^{32}$ as
``conservative"; CDF and D0 are near meltdown as these rates are
approached, and such multi-megahertz rates even in fixed-target mode
are not at all commonplace. Nevertheless it is an accurate expression
of conventional wisdom.

The trigger algorithms of the future are again of unprecedented
sophistication. The ``track triggers" of CDF, already being prepared
for the next run \cite{ref66}, are designed to partially reconstruct at
Level II $B\rightarrow h^+ h^-$. This was not in the game plan ten
years ago at all. And the successes of ring-imaging Cerenkov counters
make their introduction into hadron collider detectors most welcome:
Cerenkov identification has been completely absent from all such
detectors since the ISR days, and it is time to bring it back.

There is also an impressive array of new detectors. For the
electron-positron machines there is the BBC trio \cite{ref67}: BABAR,
BELLE, and CLEO. CDF and D0 will have powerful upgrades for their next
run. And, there will be novel devices such as HERA-B \cite{ref68}. Ten
years ago \cite{ref1} I looked toward a sophisticated high-rate fixed
target experiment as the future of the field. While it had the
disadvantage of low signal/background, it did not suffer the
collider-experiment disadvantage of having a circulating beam and
beampipe in the middle of the detector. HERA-B is the closest thing
being built to what I had in mind, although it manages to synthesize the
disadvantages of both approaches. But even so, I am an enthusiast: it
is an adventurous, powerful approach, and one which will face directly
the problems to be faced later at the LHC with LHC-B, or at the
TeVatron with BTeV.

In addition to these collider initiatives, there is the innovative
initiative E871 at Fermilab \cite{ref56} to look for CP violation in
hyperon decays. It again pushes the high-rate frontier in interesting
ways. And finally there are the experiments underway to measure
$\epsilon^\prime/\epsilon$, sure to take their place in the history of
the field as classics of experimental technique \cite{ref54,ref56}.

So with this array of opportunities, it seems clear that if the
standard-model CKM scenario for CP violation is correct, the present
program should provide a high-precision, quite complete description. On
the other hand, the next 20 years may be less programmatic than the
last 20. And if the picture is not correct, then the new facilities
should have enough flexibility to respond to an unpredictable change
in emphasis.

\section{Conclusion}

The progress in this field has been extraordinary, and the prospects
are extremely bright. It is due to the efforts of everyone in the
CP-business that this has come about. I offer my most warm and sincere
congratulations.

\section*{Acknowledgments}
I thank the many participants who have helped to minimize the inaccuracies
in this summary. Also on behalf of all participants I thank Tom Browder,
Sandip Pakvasa, and their excellent staff for making this meeting so
pleasant and successful. In particular, the organization of the meeting by
subject-matter is not an easy task, requiring a delicate touch, and this
has been accomplished in a splendid manner.


\section*{References}

\begin{thebibliography}{99}

\bibitem{ref1}
J. Bjorken, {\em Proceedings of the 5th Moriond Workshop},
``Flavor Mixing and CP Violation,'' La Plagne, France, January 1985, ed.
J. Tran Thanh Van (Editions Frontieres, Singapore, 1985), p. 455.

\bibitem{ref2}
N. Isgur and M. Wise, Phys. Lett. {\bf B232}, 113 (1989);
{\bf B237}, 527 (1990).

\bibitem{ref3}
In my talk at Moriond (Ref. 1), some time was devoted to
exhibiting the rough specifications of what was required.

\bibitem{ref4}

See, for example, the report of the Fermilab Physics Advisory Committee
on its specifications for the future use of the C\O\  collision region
in the \break 
Fermilab TeVatron:
http://www.fnal.gov/directorate/\hfill\break program\_planning/phys\_adv\_com/%
Aspen96Web.html

\bibitem{ref5}
P. Frampton, these proceedings.

\bibitem{ref6}
R. Forty, these proceedings.

\bibitem{ref7}
L. Wolfenstein, these proceedings.

\bibitem{ref8}
H. Fritzsch, hep-ph/9605388.

\bibitem{ref9}
B. Stech, Phys. Lett. {\bf 130B}, 189 (1983).

\bibitem{ref10}
This in particular implies, contrary to the hypothesis of ``Fritzsch
texture,'' that there is not much renormalization of the strange-quark
mass. I have been involved in Stech's hypothesis for quite some time
\hfill\break
(J. Bjorken, {\em Proceedings of the Eighteenth SLAC Summer Institute
on Particle Physics}, July, 1990, SLAC--REPORT--378, ed. J. Hawthorne
(SLAC), p. 167, and occasional seminars), and I still work on it.

\bibitem{ref11}
M. Neubert, these proceedings.  See also M. Neubert, Phys. Rev. {\bf
245}, 259 (1994).

\bibitem{ref12}
I. Dunietz, these proceedings.  See also I. Dunietz \etal,
hep-ph/9612421. 

\bibitem{ref13}
A. Kagan, these proceedings.

\bibitem{ref57}
H. Yamamoto, these proceedings.

\bibitem{ref16}
See, for example, M. Ciuchini \etal, Phys. Lett. {\bf B334}, 137
(1994),
W.~F.~Palmer and B. Stech, Phys. Rev. {\bf D48}, 4174 (1993).

\bibitem{ref19}
B. Behrens, these proceedings.

\bibitem{ref20}
M. Bauer, B. Stech, and M. Wirbel, Z. Phys. {\bf C34}, 103 (1987).

\bibitem{refaaa}
Since this meeting, responses from theorists have already appeared:
Dr. Atwood and A. Soni, hep-ph/9704357; I. Halperin and A. Zhitnitsky,
hep-ph/9704412; Wei-Shu Hou and B. Tseng, hep-ph/9705304.

\bibitem{ref21}
J. Richman in {\em Proceedings of ICHEP96}, Warsaw, Poland,
July 1996.

\bibitem{ref43}
T. Junk, these proceedings; see also Ref. 22.

\bibitem{ref15}
J. Rodriguez, these proceedings.

\bibitem{ref22}
B. Stech, these proceedings.

\bibitem{ref23}
J. Alexander, these proceedings.

\bibitem{ref24}
A. Falk, these proceeedings.

\bibitem{ref25}
M. Feindt, these proceeedings.

\bibitem{ref26}
M. Olsson, these proceedings.

\bibitem{ref27}
M. Luke, these proceedings.

\bibitem{ref17}
K. Kreuter, these proceedings.

\bibitem{ref28}
A. Ryd, these proceedings.

\bibitem{ref29}
N. G. Deshpande, these proceedings.  See also 
N. G. Deshpande and X.-G. He, Phys. Rev. Lett. {\bf 74}, 26 (1995);
M. Gronau \etal, Phys. Rev. {\bf D52}, 6374 (1995);
M. Gronau and D. London, Phys. Rev. Lett. {\bf 65}, 3381 (1990).

\bibitem{ref31}
G. Hanson \etal, Phys. Rev. Lett. {\bf 35}, 1609 (1975). 

\bibitem{ref32}
E. Poggio, H. Quinn, and S. Weinberg, Phys. Rev. {\bf D13}, 1958
(1976).

\bibitem{ref33}
J. Bjorken, Nucl. Phys. {\bf B11} (Proc. Suppl), 325 (1989).

\bibitem{ref34}
See for example A. Manohar, {\em Proceedings of the Tenth Lake Louise
Winter Institute}, ``Quarks and Colliders,'' February, 1995, ed.
A. Ashbury \etal\ (World Scientific, 1995), p. 274. 

\bibitem{ref35}
J. Gasser and H. Leutwyler, Nucl. Phys. {\bf B250}, 465 (1985).

\bibitem{ref36}
T. Schafer and E. Shuryak, hep-ph/9610451 and references therein. 

\bibitem{ref37}
D. Diakonov, nucl-th/9603023 and references therein.

\bibitem{ref38}
O. Eboli, E. Gregores, and F. Halzen, hep-ph/9611258 and
references therein. 

\bibitem{ref39}
W. Buchmuller and A. Hebecker, Phys. Letts. {\bf B355}, 573 (1995).

\bibitem{ref61}
R. Fleischer, these proceedings.

\bibitem{ref42}
J. Lewis, these proceedings.

\bibitem{ref44}
J. Ernst, these proceedings.  See also M. S. Alam \etal\
(CLEO Collaboration), Phys. Rev. Lett. {\bf 74}, 2885 (1995).

\bibitem{ref45}
C. Greub, these proceedings, and C. Greub, T. Hurth and D. Wyler,
Phys. Lett. {\bf B380}, 385 (1996); Phys. Rev. {\bf D54}, 3350 (1996);
K. G. Chetyrkin, M. Misiak and M. M\"unz, hep-ph/9612313; K. Adel
and Y.-P. Yao, Phys. Rev. {\bf D49}, 4945 (1994).

\bibitem{ref46}
R. Poling, these proceedings.

\bibitem{ref47}
G. Burdman, these proceedings; also G. Burdman and J. Kambor, Phys.
Rev. {\bf D55}, 2817 (1997).

\bibitem{ref48}
A. Ouraou, these proceedings.

\bibitem{ref50}
M. Purohit, these proceedings.
See also E. M. Aitala \etal, (FNAL  E791), Phys. Rev. Lett. {\bf 77},
2384 (1996).

\bibitem{ref51}
E. Golowich, these proceedings.

\bibitem{ref52}
Z.-Z. Xing, these proceedings.

\bibitem{ref53}
S. Hashimoto, these proceedings.

\bibitem{ref54}
T. Yamanaka, these proceedings.

\bibitem{ref55}
A. Schopper, these proceedings.

\bibitem{ref56}
M. Calvetti, these proceedings.

\bibitem{ref58}
X.-G. He, these proceedings; T. Browder {\em et al.}, hep-ph/9705320.

\bibitem{ref59}
M. Worah, these proceedings.

\bibitem{ref60}
D. Atwood, these proceedings; see also I. Dunietz,
``B Decays'' (ed. \hbox{S.~Stone),}  World Scientific (1994), p. 550. 

\bibitem{ref62}
S. Kurokawa, these proceedings.

\bibitem{ref63}
B. Cox, these proceedings.

\bibitem{ref64}
S. Stone, these proceedings.

\bibitem{ref65}
H. Kagan, these proceedings.

\bibitem{ref66}
A. Maciel, these proceedings.

\bibitem{ref67}
P. Harrison, these proceedings.

\bibitem{ref68}
K. Ehret, these proceedings.

\end{thebibliography}
\end{document}